\documentclass[prb,twocolumn,showpacs]{revtex4}
\usepackage{graphicx}

\begin{document}
\title{ Small-polaron hopping conductivity in bilayer manganite 
La$_{1.2}$Sr$_{1.8}$Mn$_{2}$O$_{7}$ }
\author{X. J. Chen, C. L. Zhang, and C. C. Almasan}
\affiliation{Department of Physics, Kent State University, Kent, Ohio 44242 }
\author{J. S. Gardner}
\affiliation{NRC-NPMR, Chalk River Laboratory, Chalk River ON KOJ 1PO, Canada}
\author{J. L. Sarrao}
\affiliation{Los Alamos National Laboratory, Los Alamos, New Mexico 87545 }
\date{\today}

\begin{abstract}
We report anisotropic resistivity measurements on a La$_{1.2}$Sr$_{1.8}$Mn$_{2}$O$_{7}$ 
single crystal over a temperature $T$ range from 2 to 400 K and in magnetic fields $H$ 
up to 14 T. For $T\geq 218$ K, the temperature dependence of the zero-field in-plane 
$\rho_{ab}(T)$ resistivity obeys the adiabatic small polaron hopping mechanism, while 
the out-of-plane $\rho_{c}(T)$ resistivity can be ascribed by an Arrhenius law with the 
same activation energy. Considering the magnetic character of the polarons and the close 
correlation between the resistivity and magnetization, we developed a model which allows 
the determination of $\rho_{ab,c}(H,T)$. The excellent agreement of the calculations with 
the measurements indicates that small polarons play an essential role in the electrical 
transport properties in the paramagnetic phase of bilayer manganites.
\end{abstract}
\maketitle

Elucidating the nature of the paramagnetic-insulating state is crucial to understanding the
correlation between the electrical transport and magnetic properties of $3d$ transition-metal 
manganese-oxides. Most previous studies of the manganite perovskites 
R$_{1-x}$A$_{x}$MnO$_{3}$ films (R$=$rare-earth ion and A$=$divalent ion) reveal that the 
high temperature $T$ resistivity follows the adiabatic small polaron transport.\cite{worl,tere} 
The effect of an applied magnetic field $H$ on the resistivity and thermal expansion above 
the Curie temperature $T_{c}$ indicates that the polarons have magnetic character.\cite{dete} 
The existence of polarons in the paramagnetic phase of bilayer manganites 
La$_{2-2x}$Sr$_{1+2x}$Mn$_{2}$O$_{7}$ ($x=0.4$) has been supported by measurements 
of Raman spectra,\cite{rome} x-ray and neutron scattering,\cite{vasi} optical conductivity 
spectra,\cite{ishi,hjlee} and thermoelectric power.\cite{zhou} However, there are no 
magneto-transport measurements which support the presence of polarons in the paramagnetic 
state of these materials.  

Recently, bilayer manganites La$_{2-2x}$Sr$_{1+2x}$Mn$_{2}$O$_{7}$ have attracted
considerable attention since: (i) the physical properties along the $ab$ plane and $c$ 
axis are strongly anisotropic, which should yield important insight into the colossal 
magnetoresistance (CMR) effect, (ii) they can be viewed as an infinite array of 
ferromagnetic metal (FM)-insulator (I)-FM junctions,\cite{kimura2} (iii) both the in-plane 
and out-of-plane magnetoresistivities are sensitive to even small magnetic fields, 
\cite{morit} pointing to their possible device applications, (iv) they display a rich 
magnetic phase diagram which depends strongly on the doping level $x$,\cite{hirota} and 
(v) they are good candidates for systematic investigations of the electrical resistivity 
in the paramagnetic regime over a broad $T$ range due to their relative lower $T_{c}$ 
compared to the manganite perovskites.

The understanding of electrical transport in the paramagnetic state and in the presence of 
an applied magnetic field and of the enhanced CMR effect in bilayer manganites is still 
incomplete and challenging. It has been found that the resistivity is 
semiconducting-like in the high $T$ paramagnetic state. On cooling, it reaches a 
maximum followed by a metallic behavior. When an external magnetic field is applied, this 
metal-insulator transition shifts to higher temperatures, the ferromagnetic transition 
broadens significantly, and a large reduction of electrical resistivity appears. It is highly 
desirable to understand the mechanism responsible for this charge dissipation and to develop 
a quantitative description of these behaviors. This is also essential to the understanding of 
the microscopic origin for the CMR effect. 

In this paper we address the above issues through magnetotransport measurements in a 
La$_{1.2}$Sr$_{1.8}$Mn$_{2}$O$_{7}$ single crystal. Our data show that the adiabatic 
small polaron hopping dominates the electrical transport of this bilayer manganite. 
Specifically, all the main characteristics of charge transport, $i.e.$, the $T$ and 
$H$ dependence of both the in-plane $\rho_{ab}$ and out-of-plane $\rho_{c}$
resistivities, the resistivity cusp, its shift to higher $T$ with increasing $H$, and the 
decrease of the resistivity with increasing $H$, are extremely well reproduced by our 
analysis based on the small polaron hopping, the existence of ferromagnetic clusters in 
the paramagnetic phase, and the close correlation between the resistivity and magnetization.

Single crystals of La$_{1.2}$Sr$_{1.8}$Mn$_{2}$O$_{7}$ were melt grown in a floating-zone 
optical image furnace in flowing oxygen. The crystal used here was cleaved from a boule 
that was grown at a rate of 4 mm/h and had the lowest impurity phase content.\cite{moreno} 
We used a multiterminal lead configuration \cite{jiang} for the simultaneous measurement 
of $\rho_{ab}$ and $\rho_{c}$ on the same single crystal, over temperatures from 2 to 400 K
and magnetic fields up to 14 T applied along the $ab$ planes. The electrical current was 
always applied along one of the crystal faces, while the top and bottom face voltages were 
measured simultaneously. 

Figure 1 shows $\rho_{ab}(T)$ and $\rho_{c}(T)$ measured in zero field and in several 
magnetic fields up to 14 T. The trends followed by these data are in good agreement with 
previous reports,\cite{morit} though our single crystal has lower resistivities. The
metal-insulator transition temperature $T_{MI}=130$ K is found to be slightly higher 
than $T_{c}=125$ K.\cite{moreno} We also found $T_{MI}^{ab}\simeq T_{MI}^{c}$ in 
the magnetic fields studied. Both $\rho_{ab}$ and $\rho_{c}$ decrease with increasing 
$H$, the cusp becomes less pronounced, and $T_{MI}$ shifts to higher temperatures. 
Recently, we found that both $\rho_{ab}$ and $\rho_{c}$ follow a $T^{9/2}$ dependence 
in the metallic regime (50 K$\leq T \leq$ 110 K) and both the in-plane $\sigma_{ab}$ 
and out-of-plane $\sigma_{c}$ conductivities obey a $T^{1/2}$ dependence at even lower 
$T$ ($T<50$ K), which are consistent with the two-magnon scattering and weak localization 
effect, respectively.\cite{zhang}

\begin{figure}[t]
\begin{center}
\includegraphics[width=\columnwidth]{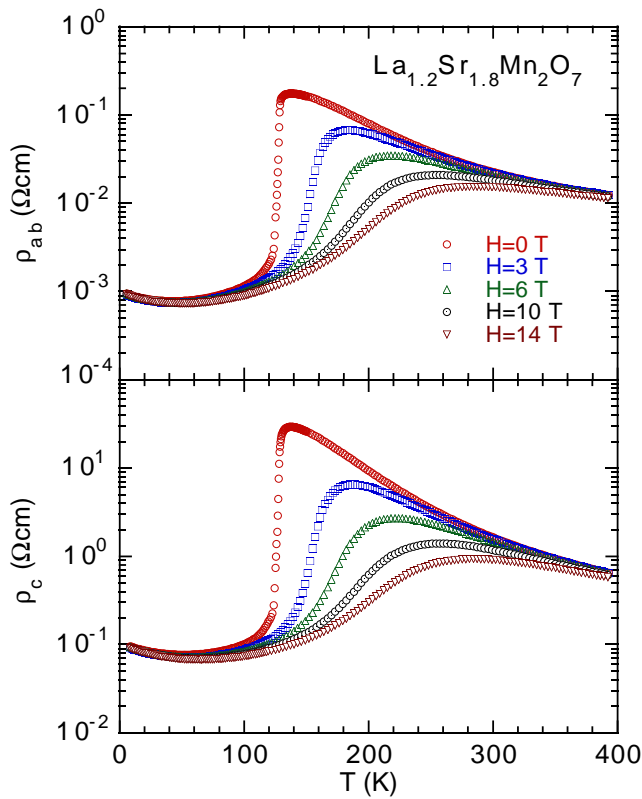}
\end{center}
\caption{In-plane $\rho_{ab}$ and out-of-plane $\rho_{c}$ resistivities as a function of 
temperature $T$ of a La$_{1.2}$Sr$_{1.8}$Mn$_{2}$O$_{7}$ single crystal for various magnetic 
fields. }
\end{figure}

The resistivity as a result of hopping of adiabatic small polarons is given by\cite{emin}
\begin{equation}
\rho=CT\exp\left(\frac{E_{A}}{k_{B}T}\right)~~.
\end{equation}
Here $k_{B}$ is Boltzmann's constant and $E_{A}$ is the activation energy. In the adiabatic 
limit, the electron motion is assumed to be much faster than the ionic motion of the lattice. 
In the approximation that all correlations except on-site Coulomb repulsion are ignored, one 
can express the prefactor $C$ as \cite{worl,raff}
\begin{equation}
C=\frac{k_{B}\Omega}{x(1-x)e^{2}a^{2}\nu}~~.
\end{equation}
Above $\Omega$ is the unit-cell volume, $x$ is the fraction concentration of occupied sites, 
$a$ is the site to site hopping distance, and $\nu$ is the frequency of the longitudinal 
optical phonons.

\begin{figure}[b]
\begin{center}
\includegraphics[width=\columnwidth]{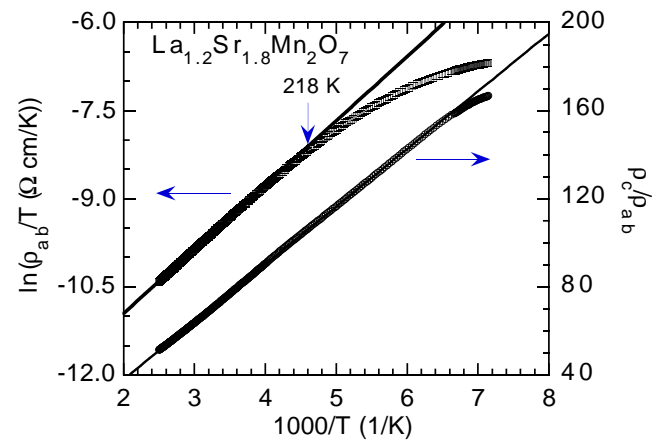}
\end{center}
\caption{Plot of $\ln(\rho_{ab}/T)$ and resistive anisotropy $\rho_c/ \rho_{ab}$, measured in 
zero field, vs $1000/T$ for a La$_{1.2}$Sr$_{1.8}$Mn$_{2}$O$_{7}$ single crystal. }
\end{figure}

To examine the polaronic nature of the high-temperature resistivity of
La$_{1.2}$Sr$_{1.8}$Mn$_{2}$O$_{7}$, we plot in Fig. 2, $\ln(\rho_{ab}/T)$ vs $1000/T$ for 
$T\geq140$ K and in zero field. The adiabatic polaron model of Eq. (1) gives a convincing 
fit to the in-plane resistivity data for $T\geq 218$ K, with a zero-field activation
energy $E_{A}^{0}=93.8$ meV and a prefactor $C=2.0\times 10^{-6}$ $\Omega$ cm/K. The fact 
that Eq. (1) is valid for $T>\Theta_{D}/2$ indicates that the Debye temperature $\Theta_{D} 
\approx 430$ K in the present bilayer compound. Indeed, recent specific heat measurements 
have shown $\Theta_{D}=425$ K in this compound.\cite{osero} The experimentally determined 
$E_{A}^{0}$ of 93.8 meV from the above resistivity data is much larger the activation energy 
$E_{S}$ of 18 meV from thermoelectric power measurements.\cite{naka} This large difference 
comes from the thermally activated nature of the hopping transport at high temperatures and 
is a characteristic signature of polaronic transport. 

Based on the experimentally determined prefactor $C$ along with the doping level $x=0.4$ 
and the lattice parameters $a=3.87\times 10^{-8}$ cm and $c=2.0\times 10^{-7}$ cm taken 
from neutron diffraction measurements,\cite{jfmitc} we estimated a characteristic frequency 
$\nu=2.24\times 10^{14}$ Hz for La$_{1.2}$Sr$_{1.8}$Mn$_{2}$O$_{7}$ by using Eq. (2). This 
value is in good agreement with the frequencies of phonon peaks in optical conductivity spectra,
\cite{ishi,hjlee} which  provides strong evidence in favor of small polaronic transport in 
the $ab$ plane of bilayer manganites.  

Figure 2 shows also the plot of the resistive anisotropy $\rho_{c}/\rho_{ab}$ vs $1000/T$ for
La$_{1.2}$Sr$_{1.8}$Mn$_{2}$O$_{7}$ for $T\geq 140$ K and zero field. Note that, for $150\leq 
T \leq 400$ K in zero field, there is the following relationship between resistivities: 
$\gamma \equiv \rho_{c}/\rho_{ab} = A+B/T$, with $A=-14.8$ and $B=2.63\times 10^{4}$ K. Since 
$\rho_{ab}(T)$ is well described by Eq. (1) for $T\geq 218$ K, it follows that $\rho_{c}(T)$
is described by an Arrhenius-type behavior with the same activation energy $E_{A}$ as 
$\rho_{ab}(T)$, if the preexponential factor $\gamma$ is taken into account. Hence,
\begin{equation}
\rho_{c}=C\gamma T\exp\left(\frac{E_{A}}{k_{B}T}\right)~~.
\end{equation}

\begin{figure}[b]
\begin{center}
\includegraphics[width=\columnwidth]{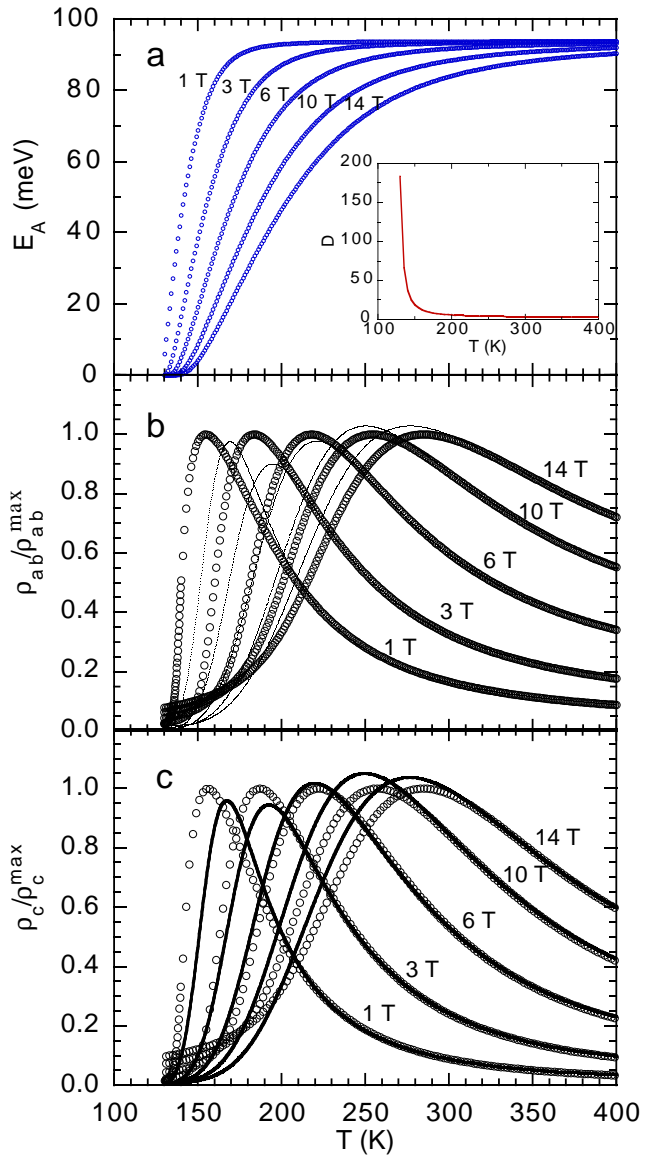}
\end{center}
\caption{ Temperature $T$ dependence of (a) the activation energy $E_{A}$, and calculated 
(solid lines) and measured (open circles) (b) normalized in-plane resistivity $\rho_{ab}$ and 
(c) normalized out-of-plane resistivity $\rho_{c}$ in a La$_{1.2}$Sr$_{1.8}$Mn$_{2}$O$_{7}$ 
single crystal for various magnetic fields $H$. Inset: Value of mean number of spins per 
cluster $D$ vs $T$. }
\end{figure}

In the case of magnetic polarons, there is a magnetic exchange contribution to the activation
energy. In the presence of a magnetic field, the activation energy in Eqs. (1) and (3) has to 
be replaced by
\begin{equation}
E_{A}=E_{A}^{0}(1-<\cos\theta_{ij}>)~~,
\end{equation}
where $\theta_{ij}$ is the angle between the spins of two Mn ion cores between which the 
$e_{g}$ electron hops. If the azimuthal angle $\phi_{i}$ is randomly distributed and if 
$\theta_{i}$, the angle the spins make with the applied field, is uncorrelated, then, by 
averaging over $\phi_{i}$, it can be shown that $<\cos\theta_{ij}>=<\cos \theta_{i}>^{2}$. 
\cite{viret} The local magnetization $M$ can also be expressed as a function of 
$\theta_{i}$, $i.e.$, $M=M_{s}<\cos\theta_{i}>$, where $M_{s}$ is the saturation 
magnetization. Then, Eq. (4) becomes
\begin{equation}
E_{A}=E_{A}^{0}\left[1-(\frac{M}{M_{s}})^{2}\right]~~.
\end{equation}
This equation shows that the magnetic field affects the activation energy through the 
magnetization. At present, there is not an agreement on the theories proposed responsible 
for the magnetic properties of manganites. It has been shown \cite{viret,bril,dion} that 
the Brillouin function $B_{s}(\lambda)$ approximately provides a quantitative description 
of the reduced magnetization $M/M_{s}$ observed experimentally. It is therefore reasonable 
to take $M/M_{s}\simeq B_{s}(\lambda)$ with 
\begin{equation}
B_{s}(\lambda)=\frac{2S+1}{2S}\coth\left(\frac{2S+1}{2S}\lambda\right)
-\frac{1}{2S}\coth\left(\frac{1}{2S}\lambda\right).
\end{equation}
Here $S$ is the average spin and the exchange coefficient and varies with doping as 
$S=3/2+(1-x)/2$. An empirical model \cite{dion} is used to sort out the magnetic field and 
temperature dependence of magnetization via the self-consistent equation 
\begin{equation}
\lambda=\frac{\mu H}{k_{B}T}+3\frac{S}{S+1}\frac{T_{c}}{T}\frac{M}{M_{s}}~~,
\end{equation}
where the effective magnetic moment $\mu/\mu_{B}=gS$ with $\mu_{B}$ being the Bohr magneton 
and $g$ being the gyromagnetic ratio. 

We note that, when using the mean-field expression for $M/M_{s}$ to analyze their measured 
magnetization data of pseudocubic manganese-oxide perovskites, Sun $et$ $al.$ \cite{jzsun}
found that $\mu/\mu_{B}=gS$ should be replaced by $\mu/\mu_{B}=DgS$, where $D$ is the mean 
number of spins per cluster. The ferromagnetic character in the paramagnetic phase of 
La$_{1.2}$Sr$_{1.8}$Mn$_{2}$O$_{7}$ has already been revealed by magnetization measurements 
\cite{moreno,pott} and other experiments.\cite{perr} Moreover, it was suggested \cite{dete} 
that the size of the ferromagnetic clusters correlates with the magnetic correlation length 
$\xi$, which increases slightly with decreasing temperature from the high-temperature 
paramagnetic side and suddenly diverges at $T_{c}$.\cite{dete,sro3} Thus the temperature
dependence of $D$ should reflect this behavior. We determined $D(T)$ values from the 
isothermal resistivity curves. The temperature dependence of $D(T)$ can be expressed as 
$D=2.7+2^{-3/2}csch ^{3/2}((T-T_{c})/320)$, which is shown in the inset of Fig. 3(a). 

We calculated the $T$ and $H$ dependences of $E_{A}$ and $\rho_{ab,c}$ from Eqs. (5)-(7) 
and from Eq. (1). $\rho_{c}(H,T)$ was then determined from the experimentally measured 
anisotropy $\gamma (H,T)$. These results for magnetic fields of 1, 3, 6, 10, and 14  T are 
shown in Figs. 3(a)--3(c) along with the experimental data of resistivities for comparison. 
In these calculations, we took $g=2$, $S=1.8$ (valid for $x=0.4$), $T_{c}=125$ K, $E_{A}^{0} 
= 93.8$ meV, and $C=2.0\times 10^{-6}$ $\Omega$ cm/K. There is a good agreement between the 
calculated and experimental results in the paramagnetic state.

As Fig. 3(a) shows, $E_{A}$ decreases slowly with decreasing $T$ and suddenly drops near 
$T_{MI}$. This characteristic behavior is responsible for the resistivity cusp shown in 
Figs. 3(b) and 3(c). With increasing magnetic field, $E_{A}$ is suppressed and its onset 
shifts systematically to high temperatures. This is the origin of the decrease of the 
resistivity as well as the shift of the resistivity cusp to higher $T$ with increasing $H$.  

In summary, the small polaron model and the existence of the ferromagnetic clusters in the 
paramagnetic phase reproduce extremely well the $T$ and $H$ dependence of $\rho_{ab,c}$ for 
bilayer manganite La$_{1.2}$Sr$_{1.8}$Mn$_{2}$O$_{7}$. Moreover, the present model also 
accounts for the resistivity cusp, its shift to higher $T$ with increasing $H$, and the 
decrease of the resistivity with increasing $H$. Hence, this work provides direct evidence 
of the presence of adiabatic small polarons in bilayer manganites and their essential role 
in both the electrical transport and the CMR effect.

This research was supported at KSU by the National Science Foundation under Grant No. 
DMR-0102415. The work at LANL was performed under the auspices of the U.S. Department of 
Energy.

\end{document}